\documentclass[10pt, final, conference, letterpaper, onecolumn, twoside, romanappendices]{./IEEEtran}

\usepackage{cite}
\usepackage{graphicx}
\usepackage{subfigure}
\usepackage{amsmath}
\usepackage{amssymb}
\usepackage{mathtools}
\usepackage{bm}
\usepackage{epsfig}
\usepackage{times}
\usepackage{color}
\usepackage{url}
\usepackage{array}
\usepackage{multirow}
\usepackage{rotating}
\usepackage{extarrows}
\usepackage{mathabx}
\usepackage{wasysym}

\newtheorem{pavikl}{\textbf{Lemma}}

\newtheorem{pavike}{\textbf{Example}}

\newcommand{\argmax}{\operatornamewithlimits{argmax}}

\title{Network Simplification for Secure AF Relaying}%
\author{\IEEEauthorblockN{Tulika Agrawal and Samar Agnihotri}%
\IEEEauthorblockA{School of Computing and Electrical Engineering, Indian Institute of Technology Mandi, HP - 175$\,$001, India}%
Email: tulika\_agrawal@students.iitmandi.ac.in, samar@iitmandi.ac.in%
}

\begin{document}

\maketitle

\begin{abstract}
We consider a class of Gaussian layered networks where a source communicates with a destination through $L$ intermediate relay layers with $N$ nodes in each layer in the presence of a single eavesdropper which can overhear the transmissions of the nodes in the last layer. For such networks we address the question: what fraction of maximum secure achievable rate can be maintained if only a fraction of available relay nodes are used in each layer?  In particular, we provide upper bounds on additive and multiplicative gaps between the optimal secure AF when all $N$ relays in each layer are used and when only $k, 1 \le k < N$, relays are used in each layer. We show that asymptotically (in source power), the additive gap increases at most logarithmically with ratio $N/k$ and $L$, and the corresponding multiplicative gap increases at most quadratically with ratio $N/k$ and $L$. To the best of our knowledge, this work offers the first characterization of the performance of network simplification in layered amplify-and-forward relay networks in the presence of an eavesdropper.
\end{abstract}


\section{Introduction}
\label{sec:Intro}
\IEEEPARstart{T}{he} inherent broadcast nature of wireless channel makes wireless transmission susceptible to eavesdropping within the communication range of the source. Traditionally security in wireless networks has mainly been considered at higher layers using cryptographic methods. However, recent advances in computation technology pose serious threats to such frameworks, motivating researchers to explore alternative security solutions which offer unbreakable and quantifiable secrecy. \textit{Physical layer security} has emerged as a viable solution. The fundamental principle behind physical layer security is to exploit the inherent randomness of noise and communication channels to limit the amount of information that can be extracted at `bit' level by an unauthorized receiver. Physical layer security builds upon the pioneering results developed by Wyner \cite{075wyner} which established the possibility of achieving information-theoretic security by exploiting the noise of communication channel and for the first time showed that secure communication is possible if the eavesdropper's channel is a \textit{degraded} version of the destination channel. Later in \cite{078cheongHellman}, Wyner's result was extended to Gaussian wire-tap channels. These results are further extended to various models such as multi-antenna systems \cite{105paradaBlahut, 110khistiWornell}, multiuser scenarios \cite{108liuMaricSpasojevicYates,108khistiTchamkertenWornell}, fading channels \cite{108liangPoorShamai, 108gopalaLaiGamal}.

An interesting direction of work on secure communication in the presence of eavesdropper(s) is one in which the source communicates with the destination via relay nodes  \cite{108laiGamal, 110dongHanPetropuluPoor, 110zhangGursoy, 110zhangGursoy2, 113yangLiMaChing}. In the existing literature such work has been considered in one or both of the following two scenarios: (A) relay nodes employ relaying schemes, such as Decode-and-Forward (DF), (B) the source communicates with the destination over a two-hop relay network. We argue that computationally complex relaying schemes such as DF render end-to-end performance characterization intractable, without providing any performance guarantee in general channel and network scenarios. Therefore, we consider one of the simplest relaying scheme: Amplify-and-Forward (AF) that allows us to provide guarantees on the optimal performance over a wide range of channel conditions. A node performing AF-relaying scales and forwards the signals received at its input. Further, with AF-relaying the end-to-end performance characterization problem remains tractable for much larger class of networks than those that can be considered with other relaying schemes.

The characterization of the optimal secure AF rate is a computationally hard problem for general layered network. Thus, in this paper we introduce an approach based on network simplification to reduce the computational effort of approximating the maximum achievable secure AF rate in general layered networks. Consider a network scenario where source $s$ is connected to a destination $t$ via a network of wireless relays in the presence of an eavesdropper. The eavesdropper can overhear the transmissions of a subset of relay nodes in the network. To characterize the  maximum secure AF rate one has to optimize it over the scaling factors of all relay nodes. However, there can be several relay nodes which contribute marginally to the achievable secrecy rate. Shutting down those relays saves physical resources without compromising much on the performance. At the same time, computational effort of calculating optimal secure AF rate is reduced greatly as now one needs to optimize the secrecy rate over scaling factors of fewer relay nodes. In this paper we aim to understand what portion of the maximum achievable secrecy rate can be maintained if only a fraction of available relay nodes are used. 

Previous work on network simplification embraces two major threads: one pertaining to the characterization of the fraction of achievable rate/capacity that can be maintained with a selected subset of available relay nodes and the other pertaining to the design of efficient algorithms for the selection of a subset of the best relay nodes. In the first direction, for the Gaussian N-relay diamond network, \cite{114nazarogluFragouli} characterizes the fraction of the capacity when $k$ out of $N$ available relay nodes are used. In \cite{111nazarogluFragouli}, the work of  \cite{114nazarogluFragouli} were extended to diamond network with multiple antennas at the source and the destination for some scenarios. Authors in \cite{112agnihotriJaggiChen3} provide upper bounds on multiplicative and additive gaps between optimal AF rates with and without network simplification for the Gaussian diamond network and a class of symmetric layered networks. Recent work in \cite{116ezzeldinSenguptaFragouli}, characterizes the guarantees achievable over arbitrary layered Gaussian networks, however restricted to the selection of exactly one relay from each layer. The performance guarantees for the scenario where a subset of two relays per layer is selected from a network with two layers of three relays each is also provided. The progress in the second direction is made by \cite{114brahmaSenguptaFragouli} and \cite{114kolteOzgur}, where low-complexity heuristic algorithms for the selection of near-optimal relay subnetwork of a given size from a layered Gaussian relay network is provided. Previously,  in cooperative communication literature also, the notion of relay selection has been used \cite{106bletasKhistiReedLippman,107zhaoAdveLim,108caiShenMarkAlfa}. However, such prior work used relay subnetwork selection in a restricted sense (selecting the best single relay node among N relays in one layer networks).

In the above-mentioned work the main objective was throughput maximization. To the best of our knowledge, no work has been done hitherto towards the characterization of the performance of network simplification for wireless relay networks in the presence of an eavesdropper with the objective of secrecy rate maximization which is an even harder problem. As a first step in this direction, we provide the optimal secure AF rate characterization in the communication scenarios where the source communicates with the destination over $L$ layers of relay nodes $(L \geq 1)$, in the presence of an eavesdropper which overhears the transmissions of the nodes in the last layer and any number $k$ of relay nodes in each layer are used,  $1 \le k < N$. The eavesdropper being a passive entity, a realistic eavesdropper scenario is the one where nothing about the eavesdropper's channel is known, neither its existence, nor its channel state information (CSI). However, the existing work on secrecy rate characterization assumes one of the following: (1) the transmitter has prefect knowledge of the eavesdropper channel states, (2) \textit{compound channel:} the transmitter knows that the eavesdropper channel can take values from a finite set \cite{109liangKramerPoorShamai, 109kobayashiLiangShamaiDebbah, 109ekremUlukus}, and (3) \textit{fading channel:} the transmitter only knows distribution of the eavesdropper channel \cite{108liangPoorShamai, 108gopalaLaiGamal}. In this paper, we assume that the CSI of the eavesdropper channel is known perfectly for the following two reasons. First, This provides an upper bound to the achievable AF secrecy rate for the scenarios where we have imperfect knowledge of the eavesdropper channel. For example, the lower (upper) bound on the compound channel problem can be computed by solving the perfect CSI problem with the worst (best) channel gain from the corresponding finite set. Further, this also provides a benchmark to evaluate the performance of achievability schemes in such imperfect knowledge scenarios. Second, this assumption allows us to focus on the nature of the optimal solution and information flow, instead of on complexities arising out of imperfect channel models.

\textit{Organization:} In Section~\ref{sec:sysMdl} we introduce a general wireless layered relay network model and formulate the problem of maximum achievable secrecy rate with amplify-and-forward relaying in such networks. Section~\ref{sec:NetSimDiamondNet} addresses the performance of network simplification in the Gaussian $N$-relay diamond network, layered network with $L=1$ layer and computes additive and multiplicative gaps between the maximum secure AF rates achievable when $N$ and $k\ (k < N)$, relays are used. In Section~\ref{sec:ECGALnet} we consider a class of symmetric layered networks and compute additive and multiplicative gaps between the optimal secure AF rates obtained when when $N$ and $k\ (k < N)$, relays are used. Section~\ref{sec:cnclsn} concludes the paper.

\section{System Model}
\label{sec:sysMdl}
\begin{figure}[!t]
\centering
\includegraphics[width=3.5in]{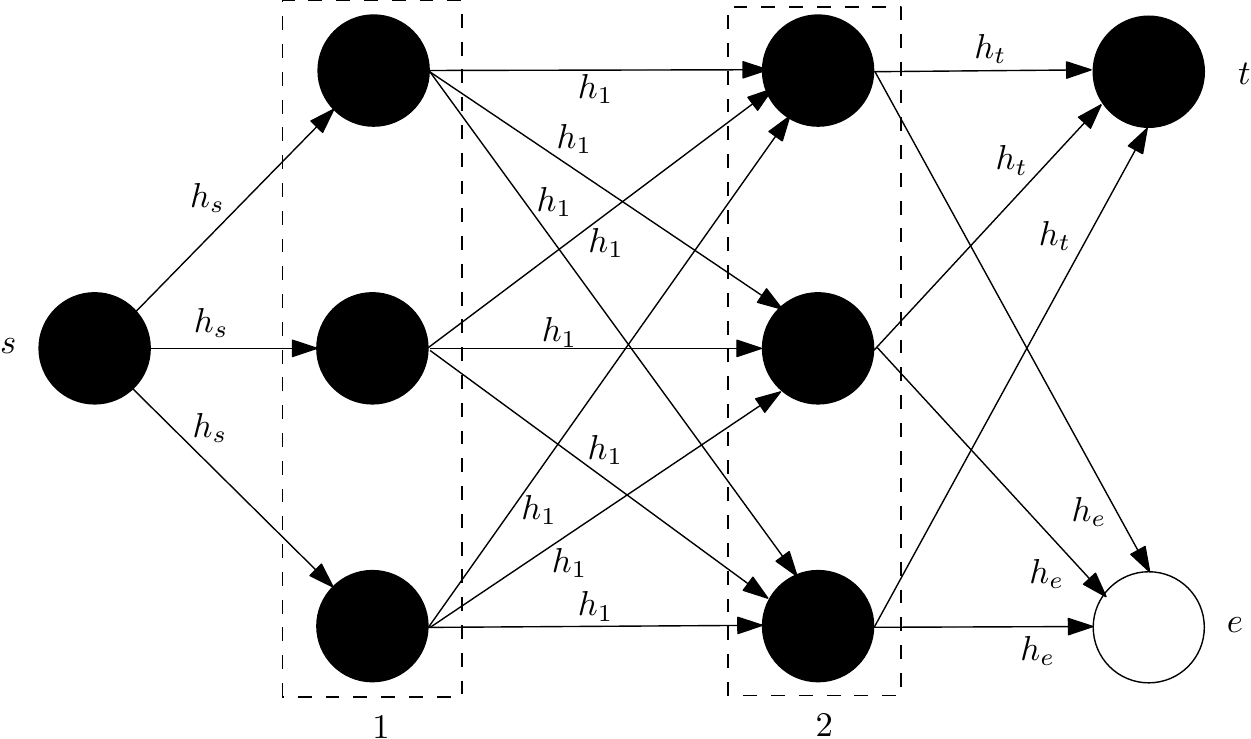}
\caption{An ECGAL network with 3 relay layers between the source $s$ and the destination $t$. Each layer contains two relay nodes. The eavesdropper overhears the transmissions from the relays in layer $2$.}
\label{fig:layrdNetExa}
\vspace{-0.2in}
\end{figure}

Consider a $(L+2)$-layer wireless network with directed links. The source $s$ is at layer `$0$', the destination $t$ is at layer `$L+1$' and the relays from the set $R$ are arranged in $L$ layers between them. The $l^{th}$ layer contains $n_l$ relay nodes, $\sum _{l-1}^{L} n_l = |R|$. The source $s$ transmits message signals to the destination $t$ via $L$ relay layers.  However, the signals transmitted by the relays in the last layer are also overheard  by the eavesdropper $e$. An instance of such a network is given in Figure~\ref{fig:layrdNetExa}. Each node is assumed to have a single antenna and operate in full-duplex mode.

At instant $n$, the channel output at node $i, i \in R \cup \{t, e\}$, is
\begin{equation}
\label{eqn:channelOut}
y_i[n] = \sum_{j \in {\mathcal N}(i)} h_{ji} x_j[n] + z_i[n], \quad - \infty < n < \infty,
\end{equation}
where $x_j[n]$ is the channel input of node $j$ in neighbor set ${\mathcal N}(i)$ of node $i$. In \eqref{eqn:channelOut}, $h_{ji}$ is a real number representing the channel gain along the link from the node $j$ to the node $i$ and constant over time (as in \cite{112agnihotriJaggiChen}, for example) and known (even for the eavesdropper channels) throughout the network as in \cite{110dongHanPetropuluPoor,110zhangGursoy,110zhangGursoy2}. All channel gains between the nodes in two adjacent layers are assumed to be equal - \textit{``Equal Channel
Gains between Adjacent Layers (ECGAL)''} networks \cite{112agnihotriJaggiChen}. Source symbols $x_s[n], - \infty < n < \infty$, are i.i.d. Gaussian random variables with zero mean and variance $P_s$ that satisfy an average source power constraint, $x_s[n] \sim {\cal N}(0, P_s)$. Further, $\{z_i[n]\}$ is a sequence (in $n$) of i.i.d. Gaussian random variables with $z_i[n] \sim {\cal N}(0, \sigma^2)$. We assume that $z_i's$ are independent of the input signal and of each other. We also assume that the $i^{\textrm{th}}$ relay's transmit power is constrained as:
\begin{equation}
\label{eqn:pwrConstraint}
E[x_i^2[n]] \le P_i, \quad - \infty < n < \infty
\end{equation}

Each relay node amplifies and forwards the noisy signal sum received at its input. More precisely, relay node $i, i \in R$, at instant $n+1$ transmits the scaled version of $y_i[n]$, its input at time instant $n$, as follows
\begin{equation}
\label{eqn:AFdef}
x_i[n+1] = \beta_i y_i[n], \quad 0 \le \beta_i^2 \le \beta_{i,max}^2 = P_i/P_{R,i},
\end{equation}
where $P_{R,i}$ is the received power at node $i$ and choice of scaling factor $\beta_i$ satisfies the power constraint \eqref{eqn:pwrConstraint}.

Assuming equal delay along each path, for the network in Figure \ref{fig:layrdNetExa}, the copies of the source signal ($x_s[.]$) and noise signals ($z_i[.]$), respectively, arrive at the destination and the eavesdropper along multiple paths of the same delay. Therefore, the signals received at the destination and eavesdropper are free from intersymbol interference (ISI). Thus, we can omit the time indices and use equations \eqref{eqn:channelOut} and \eqref{eqn:AFdef} to write the input-output channel between the source $s$ and the destination $t$ as
\begin{equation}
\label{eqn:destchnl}
y_t = \left[\sum\limits_{(i_1,...,i_L) \in K_{st}}\!\!\!\!\!\!\!\!\! h_{s,i_1}\beta_{i_1}h_{i_1, i_2}...h_{i_{L-1}, i_L}\beta_{i_L} h_{i_L, t}\right] x_s + \sum\limits_{l=1}^L \sum\limits_{j-1}^{n_l}\left[\sum\limits_{(i_1,...,i_{L-l}) \in K_{lj,t}} \!\!\!\!\!\!\!\!\!\beta_{lj} h_{lj,i_1}...\beta_{i_{L-l}} h_{i_{L-l},t}\right] z_{lj} + z_t
\end{equation}
where $K_{st}$ is the set of $L$-tuples of node indices corresponding to all paths from source $s$ to destination $t$ with path delay $L$. Similarly, $K_{lj,t}$ is the set of $L-l$- tuples of node indices corresponding to all paths from the $j^{th}$ relay of $l^{th}$ layer to the destination with path delay $L-l + 1$.

We introduce modified channel gains as follows. For all the paths between the source and the  destination:
\begin{equation}
h_{st} = \sum\limits_{(i_1,...,i_L) \in K_s} h_{s,i_1}\beta_{i_1}h_{i_1, i_2}...h_{i_{L-1}, i_L}\beta_{i_L} h_{i_L, t}
\end{equation}
For all the paths between the $j^{th}$ relay of $l^{th}$ layer to destination $t$ with path delay $L-l + 1$:
\begin{equation}
h_{lj,t} = \sum\limits_{(i_1,...,i_{L-l}) \in K_{lj}} \beta_{lj} h_{lj,i_1}...\beta_{i_{L-l}} h_{i_{L-l},t}
\end{equation}

In terms of these modified channel gains, the source-destination channel in \eqref{eqn:destchnl} can be written as: 
\begin{equation}
\label{eqn:destchnlmod}
y_t = h_{st} x_s + \sum_{l=1}^{L} \sum_{j=1}^{n_l} h_{lj,t} z_{lj} + z_t,
\end{equation}

Similarly, the input-output channel between the source and the eavesdropper can be written as 
\begin{equation}
\label{eqn:evechnlmod}
y_e = h_{se} x_s + \sum_{l=1}^{L} \sum_{j=1}^{n_l} h_{lj,e} z_{lj} + z_t,
\end{equation}

The secrecy rate at the destination for such a network model can be written as \cite{075wyner},  
$R_s(P_s)= [I(x_s;y_t)-I(x_s;y_e)]^+$,
where $I(x_s;y)$ represents the mutual information between random variable $x_s$ and $y$ and $[u]^+=\max\{u,0\}$.
The secrecy capacity is attained for the Gaussian channels with the Gaussian input $x_s \sim \mathcal{N}(0,P_s)$, where $\mathbf{E}[x_s^2] = P_s$,\cite{078cheongHellman}. Therefore, for a given network-wide scaling vector $\bm{\beta} = (\beta_{li})_{1 \le l \le L, 1 \le i \le n_l}$, the optimal secure AF rate for the channels in \eqref{eqn:destchnlmod} and \eqref{eqn:evechnlmod} can be written as the following optimization problem.
\begin{subequations}
\label{eq:optSecrate}
\begin{align}
R_s(P_s) &=  \max_{\bm{\beta}} \left[R_t(P_s,\bm{\beta}) - R_e(P_s,\boldsymbol{\beta})\right]\\
         &=  \max_{\boldsymbol{\beta}} \left[\frac{1}{2}\log\frac{1+SNR_t(P_s,\bm{\beta})}{1+SNR_e(P_s,\bm{\beta})}\right],
\end{align}  
\end{subequations}
where $SNR_t(P_s,\bm{\beta})$, the signal-to-noise ratio at the destination $t$ is:
\begin{equation}
\label{eqn:snrt}
SNR_t(P_s,\bm{\beta}) = \frac{P_s}{\sigma^2}\frac{h_{st}^2}{1 + \sum_{l=1}^{L} \sum_{j=1}^{n_l} h_{lj,t}^2}
\end{equation}
and similarly, $SNR_e(P_s,\bm{\beta})$ is
\begin{equation}
\label{eqn:snre}
SNR_e(P_s,\bm{\beta}) = \frac{P_s}{\sigma^2}\frac{h_{se}^2}{1 + \sum_{l=1}^{L} \sum_{j=1}^{n_l} h_{lj,e}^2}
\end{equation}

Given the monotonicity of the $\log(\cdot)$ function, we have
\begin{align}
\bm{\beta}_{opt} &= \argmax_{\bm{\beta}} \left[R_t(P_s,\bm{\beta}) - R_e(P_s,\boldsymbol{\beta})\right] \nonumber\\
                 &= \argmax_{\bm{\beta}} \frac{1+SNR_t(P_s,\bm{\beta})}{1+SNR_e(P_s,\bm{\beta})} \label{eqn:eqProb}
\end{align}
However, so far, there exists no closed-form expression or polynomial time algorithm to exactly solve the problem \eqref{eqn:eqProb} even for general two-hop (diamond) relay networks except for few specific cases where eavesdropper's channel is a degraded or scaled version of destination channel  \cite{115sarmaAgnihotriKuri2} and solving it for general layered network with more than one layer is an even harder problem. Therefore in this paper, based on the notion of network simplification
\cite{114nazarogluFragouli}, we introduce an approach to reduce the computational effort of solving this problem by selecting the best subset of $k_l$ relays among the set of $n_l$ relays in the $l^{th}$ layer. Thus the proposed scheme leads to an exponential reduction in the search space to solve the problem  \eqref{eqn:eqProb} without significantly compromising on the optimal performance, in terms of additive and multiplicative gaps between the corresponding performances, as we show below.
In the following, we discuss the performance of this network simplification based approach to approximate the maximum achievable secure AF rate first, for the symmetric Gaussian N-relay diamond networks and then, for a class of symmetric layered networks.

\section{Performance of Network Simplification in Diamond network}
\label{sec:NetSimDiamondNet}
Consider the symmetric diamond network, layered network with only a single layer of relays between the source and the destination, $L=1$, with $N$ relay nodes. 
Using \eqref{eqn:snrt} and \eqref{eqn:snre}, the $SNR_t$ and $SNR_e$ in this case are:
\begin{align*}
SNR_t = \frac{P_s h_s^2 }{\sigma^2} \frac{(\sum_{i=1}^{N} \beta_i )^2 h_t^2}{1 + \left(\sum_{i=1}^{N} \beta_i^2 \right) h_t^2 } &&\mbox{ and }&& SNR_e = \frac{P_s h_s^2 }{\sigma^2} \frac{(\sum_{i=1}^{N} \beta_i )^2 h_e^2}{1 + \left(\sum_{i=1}^{N} \beta_i^2 \right) h_e^2 }
\end{align*}
\begin{pavikl}
\label{lemma:diamondNetReducedBeta1}
For symmetric diamond network, $\bm{\beta}_{opt}$ in \eqref{eqn:eqProb} is:
\begin{align*}
\beta_{1,opt},\cdots,\beta_{N,opt}=\beta_{opt}=
			\begin{cases}
               \min(\beta_{max}^2, \beta_{glb}^2), \quad \mbox{if } h_{t}>h_e,\\
               0, \quad \mbox{otherwise}
            \end{cases}
\end{align*}
where
\begin{align*}
\beta_{max}^2 = \frac{P}{P_s h_s^2 + \sigma^2} & &\mbox{and} & & \beta_{glb}^2 = \sqrt{\frac{1}{N^2 h_t^2 h_e^2 \left(1+N\frac{P_s h_s^2}{\sigma^2} \right)}}
\end{align*}
\end{pavikl}
\begin{IEEEproof}
To find the value of $\bm{\beta}_{opt}$ that maximizes the corresponding secrecy rate, equating the partial derivative of the secrecy rate in \eqref{eqn:eqProb} with respect to $\beta_{i}$ to zero, we get a system of $N$ simultaneous polynomial equations. Without any loss of generality, subtracting the equation corresponding to the partial derivative with respect to $\beta_{1}$ from the rest of $N-1$ equations, it is easy to prove that $\beta_{1} = \beta_{2} = \cdots = \beta_{N} = \beta$ is the only root of this system of equations. Substituting this solution in one of the equations, we get
\begin{equation*}
\beta \left(N^2 h_t^2 h_e^2 (N P_s h_s^2 + \sigma^2)\beta^4 - \sigma^2\right) = 0
\end{equation*}
This equation has the following two distinct and real solutions for the stationary points of the secrecy rate with respect to $\beta$:
\begin{equation*}
\beta_{z} = 0, \mbox{ and } \beta_{glb}^2 = \sqrt{\frac{1}{N^2 h_t^2 h_e^2 \left(1+N\frac{P_s h_s^2}{\sigma^2} \right)}}
\end{equation*}

Now using the second derivative test, we can prove that when $h_t>h_e$, then $\beta_{z}$ and $\beta_{glb}$ are the points of global minimum and maximum, respectively. Similarly, for $he > h_t$, $\beta_{z}$ and $\beta_{glb}$ can be proved to be the points of global maximum and minimum, respectively. Given the convex nature of the secrecy rate function with respect to $\beta$ for $h_t>h_e$ and that $\beta_{max}$ is the largest value of the scaling factor $\beta$, we have the result $ \beta_{opt}^2 = \min(\beta_{max}^2, \beta_{glb}^2), \, \mbox{if } h_t > h_e$.
\end{IEEEproof}

With these optimum scaling factors, the SNR at the destination with $N$ relay nodes is given as follows
\begin{align}
SNR_{t,opt}^N &= \frac{P_s h_s^2}{\sigma^2} \frac{N^2 (\beta^N_{opt})^2 h_t^2}{1+N (\beta^N_{opt})^2 h_t^2} \label{eqn:snr_t_N_diamond}
\end{align}
We use superscript $N$ to emphasize that the optimal scaling factors are computed for all the $N$ nodes.

Similarly, the SNR at the eavesdropper with $N$ relay nodes is given as follows
\begin{align}
SNR_{e,opt}^N &= \frac{P_s h_s^2}{\sigma^2} \frac{N^2 (\beta^N_{opt})^2 h_e^2}{1+N (\beta^N_{opt})^2 h_e^2} \label{eqn:snr_e_N_diamond}
\end{align}

Now consider the network simplification scenario where only $k$ out of $N$ available relays are used, $1 \le K < N$. Using Lemma~\ref{lemma:diamondNetReducedBeta1}, we can solve secrecy rate optimization problem of \eqref{eqn:eqProb} and obtain the optimal solution $ {\beta}_{opt}^k $, given as follows:
\begin{align*}
(\beta_{opt}^k)^2 &= \min\left\{(\beta_{max})^2, (\beta_{glb}^k)^2\right\}
\end{align*}
where
\begin{align*}
(\beta_{glb}^k)^2 &= \sqrt{\frac{1}{k^2 h_t^2 h_e^2 \left(1+k\frac{P_s h_s^2}{\sigma^2} \right)}}
\end{align*}

The corresponding optimal SNR at the destination and the eavesdropper is given as:
\begin{align}
SNR_{t,opt}^k &= \frac{P_s h_s^2}{\sigma^2} \frac{k^2 (\beta^k_{opt})^2 h_t^2}{1+k(\beta^k_{opt})^2 h_t^2}\label{eqn:snr_t_k_diamond}\\
SNR_{e,opt}^k &= \frac{P_s h_s^2}{\sigma^2} \frac{k^2 (\beta^k_{opt})^2 h_e^2}{1+k(\beta^k_{opt})^2 h_e^2}\label{eqn:snr_e_k_diamond}
\end{align}

Let 
\begin{equation}
\label{eqn:Rs_N}
R_s^N = \frac{1}{2}\log\left(\frac{1+SNR_{t,opt}^N}{1+SNR_{e,opt}^N}\right) 
\end{equation}
and
\begin{equation}
\label{eqn:Rs_k}
R_s^k = \frac{1}{2}\log\left(\frac{1+SNR_{t,opt}^k}{1+SNR_{e,opt}^k}\right)
\end{equation} 
denote the optimal secure AF rate achieved by using all $N$ relays and any $k$ relays out of available $N$ relays of the diamond network, respectively. In the following we compute the additive gap $R_s^N - R_s^k$ for large $P_s$ and the multiplicative gap $R_s^N/R_s^k$ for small $P_s$.

\textbf{Case I:} $\beta_{opt}^2 = \beta_{max}^2$\\
\begin{equation}
R_s^N = \frac{1}{2} \log \left(\frac{1+\frac{P_s h_s^2}{\sigma^2} \frac{N^2 P h_t^2}{P_s h_s^2 + \sigma^2 + N P h_t^2}}{1+\frac{P_s h_s^2}{\sigma^2} \frac{N^2 P h_e^2}{P_s h_s^2 + \sigma^2 + N P h_e^2}}\right)
\end{equation}
\begin{equation}
R_s^k = \frac{1}{2} \log \left(\frac{1+\frac{P_s h_s^2}{\sigma^2} \frac{k^2 P h_t^2}{P_s h_s^2 + \sigma^2 + k P h_t^2}}{1+\frac{P_s h_s^2}{\sigma^2} \frac{k^2 P h_e^2}{P_s h_s^2 + \sigma^2 + k P h_e^2}}\right)
\end{equation}
\begin{align}
\lim_{P_s \rightarrow \infty} R_s^N - R_s ^k & = \lim_{P_s \rightarrow \infty} \frac{1}{2}\log\!\left[\frac{1+\frac{N^2 P h_t^2}{\sigma^2}}{1+\frac{N^2 P h_e^2}{\sigma^2}}\right] -\frac{1}{2}\log\!\left[\frac{1+\frac{k^2 P h_t^2}{\sigma^2}}{1+\frac{k^2 P h_e^2}{\sigma^2}}\right]\nonumber\\
		& \leq \lim_{P_s \rightarrow \infty} \frac{1}{2}\log\!\left[\left(\frac{N}{k}\right)^2 \frac{1+\frac{k^2 P h_e^2}{\sigma^2}}{1+\frac{N^2 P h_e^2}{\sigma^2}}\right]\nonumber\\
		& = \lim_{P_s \rightarrow \infty} \frac{1}{2}\log\!\left[ \frac{1+\frac{\sigma^2}{k^2 P h_e^2}}{1+\frac{\sigma^2}{N^2 P h_e^2}}\right]\nonumber\\
			&\leq \frac{1}{2}\log\!\left[1+\frac{\sigma^2}{P h_e^2}\left(\frac{1}{k^2}-\frac{1}{N^2}\right)\right]
\end{align}

\begin{align*}
\lim_{P_s \rightarrow 0} R_s^N / R_s ^k & = \lim_{P_s \rightarrow 0}  \log\left. \left[  \frac{1+\frac{P_s h_s^2}{\sigma^2}  \frac{N^2 P h_t^2}{\sigma^2+N P h_t^2}}{1+\frac{P_s h_s^2}{\sigma^2}  \frac{N^2 P h_e^2}{\sigma^2+N P h_e^2}} \right]  \middle/   \log \left[  \frac{1+\frac{P_s h_s^2}{\sigma^2}  \frac{k^2 P h_t^2}{\sigma^2+k P h_t^2}}{1+\frac{P_s h_s^2}{\sigma^2}  \frac{k^2 P h_e^2}{\sigma^2+k P h_e^2}} \right]\right.\\
		& \stackrel{(a)}{\leq} \lim_{P_s \rightarrow 0} \left(\frac{N}{k}\right)^2 \left( \frac{1+k \frac{P h_e^2}{\sigma^2}}{1+N \frac{P h_t^2}{\sigma^2}}\right) \left(\frac{1+k \frac{P h_t^2}{\sigma^2}+k^2  \frac{P_s h_s^2}{\sigma^2}\frac{P h_t^2}{\sigma^2}}{1+N \frac{P h_e^2}{\sigma^2}+N^2  \frac{P_s h_s^2}{\sigma^2}\frac{P h_e^2}{\sigma^2}}\right)\\
		& \leq \left(\frac{N}{k}\right)^2 \left( \frac{1+k \frac{P h_e^2}{\sigma^2}}{1+N \frac{P h_t^2}{\sigma^2}}\right) \frac{h_t^2}{h_e^2}\\
		& \leq \left( \frac{N}{k}\right) \left[1+ \frac{\sigma^2}{P}\left(\frac{1}{k h_e^2}-\frac{1}{N h_t^2}\right) \right]
\end{align*}
where $(a)$ follows from the fact that the arguments of $\log(.)$ in the numerator and denominator (say x and y respectively) are greater than 1 for $h_t > h_e$ (condition for non-zero secrecy rate), and under the limit $P_s \rightarrow 0 $ and comparable $h_t$ and $h_e$ the arguments are close to 1. Thus the numerator can be tightly bounded from above by $x-1$ and similarly, denominator can be tightly lower bounded by $(y-1)/y$.

\textbf{Case II:} $\beta_{opt}^2 = \beta_{glb}^2$\\
Substituting for $\beta_{opt}^N$ in \eqref{eqn:snr_t_N_diamond} and \eqref{eqn:snr_e_N_diamond} and subsequently substituting the results in \eqref{eqn:Rs_N}, we get
\begin{align}
R_s^N &= \frac{1}{2}\log\left[\frac{1+\frac{N P_s h_s^2/ \sigma^2}{1+\frac{h_e}{h_t} \sqrt{1+\frac{N P_s h_s^2}{\sigma^2}}}}{1+\frac{N P_s h_s^2/ \sigma^2}{1+\frac{h_t}{h_e} \sqrt{1+\frac{N P_s h_s^2}{\sigma^2}}}}\right]
\end{align}

Similarly,
\begin{align}
R_s^k &= \frac{1}{2}\log\left[\frac{1+\frac{k P_s h_s^2/ \sigma^2}{1+\frac{h_e}{h_t} \sqrt{1+\frac{k P_s h_s^2}{\sigma^2}}}}{1+\frac{k P_s h_s^2/ \sigma^2}{1+\frac{h_t}{h_e} \sqrt{1+\frac{k P_s h_s^2}{\sigma^2}}}}\right]
\end{align}

Thus, we have
\begin{align}
\lim_{P_s \rightarrow \infty} R_s^N - R_s ^k &= \lim_{P_s \rightarrow \infty} \frac{1}{2}\log \left[\frac{1+\frac{h_t}{h_e}\sqrt{\frac{N P_s h_s^2}{\sigma^2}}}{1+\frac{h_e}{h_t}\sqrt{\frac{N P_s h_s^2}{\sigma^2}}}\frac{1+\frac{h_e}{h_t}\sqrt{\frac{k P_s h_s^2}{\sigma^2}}}{1+\frac{h_t}{h_e}\sqrt{\frac{k P_s h_s^2}{\sigma^2}}}\right] \nonumber\\	
				&\leq \lim_{P_s \rightarrow \infty} \frac{1}{2}\log \left[\frac{1+\frac{h_t}{h_e}\sqrt{\frac{N P_s h_s^2}{\sigma^2}}}{1+\frac{h_t}{h_e}\sqrt{\frac{k P_s h_s^2}{\sigma^2}}}\right] \nonumber\\	
											   &\stackrel{(b)}{\leq} \frac{1}{4}\log\left(\frac{N}{k}\right)
\end{align}
where $(b)$ follows from the fact that ${(1+x)}/{(1+y)} \leq {x}/{y}$ for $x \geq y$.
 
And
\begin{align}
\lim_{P_s \rightarrow 0} \frac{R_s^N}{R_s^K} &= \lim_{P_s \rightarrow 0} \log \left( \left.\frac{1+\frac{N P_s h_s^2/\sigma^2}{1+{h_e}/{h_t} }}{1+\frac{N P_s h_s^2/\sigma^2}{1+{h_t}/{h_e} }}\right)\middle/\right.\log \left( \frac{1+\frac{k P_s h_s^2/\sigma^2}{1+{h_e}/{h_t} }}{1+\frac{k P_s h_s^2/\sigma^2}{1+{h_t}/{h_e} }}\right)\nonumber\\
		&\stackrel{(c)}{\le} \lim_{P_s \rightarrow 0} \left(\frac{N}{k}\right)\left(\frac{h_t + h_e + k \frac{P_s h_s^2}{\sigma^2} h_t}{h_t + h_e + N \frac{P_s h_s^2}{\sigma^2} h_e}\right) \nonumber\\
		& \leq \max\left\{ \frac{N}{k}, \frac{h_t}{h_e}\right\}
\end{align}
where $(c)$ follows from the same argument as $(a)$.
\section{Performance of Network Simplification in Layered network}
\label{sec:ECGALnet}
In this section we analyze the performance of network simplification in  ECGAL networks with two or more layers of relays between the source and the destination and each relay performing amplify-and-forward on its input signal. This is a generalization of the symmetric diamond network configuration considered in Section~\ref{sec:NetSimDiamondNet} above. For the ease of presentation, consider ECGAL networks where each layer of relay nodes has $N$ relays and all relay nodes have the same transmit power constraint $\textbf{E}X^2 \leq P$.

Using \eqref{eqn:snrt} and \eqref{eqn:snre}, the $SNR_t$ and $SNR_e$ in this case are:
\begin{align}
S\!N\!R_t &= \frac{P_s}{\sigma^2} \frac{h_s^2 H_{1,L-1}^2 \left(\sum_{n=1}^N \beta_{L,n}\right)^{2} h_t^2}{\left(\sum_{i=1}^{L-1} \!\!G_{i,L\!-\!1}^2\!\right)\!\!\left(\sum_{n=1}^N \!\!\beta_{L,n}\!\right)^{\!2}\! \! h_t^2\!+\!\!\left(\sum_{n=1}^N \beta_{L,n}^2\!\right)\!\! h_t^2+\!1}\\
S\!N\!R_e &= \frac{P_s}{\sigma^2} \frac{h_s^2 H_{1,L-1}^2 \left(\sum_{n=1}^N \beta_{L,n}\right)^{2} h_e^2}{\left(\sum_{i=1}^{L-1} \!\!G_{i,L\!-\!1}^2\!\right)\!\!\left(\sum_{n=1}^N \!\!\beta_{L,n}\!\right)^{\!2}\! \! h_e^2\!+\!\!\left(\sum_{n=1}^N \beta_{L,n}^2\!\right)\!\! h_e^2+\!1}
\end{align}
where
\begin{align*}
H_{i,j}^2 &= \prod_{k=i}^{j} \left(\sum_{n=1}^N \beta_{k,n}\right)^2 h_k^2\\
G_{i,j}^2 &= \left(\sum_{n=1}^N \beta_{i,n}^2\right) h_i^2 \prod_{k=i+1}^{j} \left(\sum_{n=1}^N \beta_{k,n}\right)^2 h_k^2
\end{align*}
The optimum solution $\bm{\beta}_{opt}$ of problem \eqref{eqn:eqProb} for an ECGAL network with $N$ relay nodes in each layer is given by the following two lemmas:
\begin{pavikl}
\label{lemma:ECGALReducedBeta1}
For a given sub-vector $(\bm{\beta}_1, \cdots, \bm{\beta}_{L-1})$, the sub-vector $\bm{\beta}_{L,opt}=(\beta_{L,1,opt},\cdots,\beta_{L,N,opt})$ of the optimum scaling factors for the nodes in the $L^{th}$ layer is:
\begin{equation*}
\beta_{L,1,opt}^2 = \ldots = \beta_{L,N,opt}^2 = \beta_{L,opt}^2 = 
                 \begin{cases}
                   \min(\beta_{L,max}^2, \beta_{L,glb}^2), \mbox{if } h_t > h_e,\\
                   0, \quad \mbox{otherwise}
                 \end{cases}
\end{equation*} 
where 
\begin{equation}
\beta_{L,glb}^2 = \frac{1}{N h_t h_e (1+N B) \sqrt{1+\frac{P_s}{\sigma^2}\frac{N A}{1+N B}}} \label{eqn:betaLglb}
\end{equation}
with $A = h_s^2 H_{1,L-1}^2 \mbox{ and } B =\sum_{i=1}^{L-1} G_{i,L-1}^2$.
\end{pavikl}
\begin{IEEEproof}

To find the value of $\bm{\beta}_{L,opt}$ that maximizes the corresponding secrecy rate, equating the partial derivative of the secrecy rate in \eqref{eqn:eqProb} with respect to $\beta_{L,n}$ to zero, we get a system of $N$ simultaneous polynomial equations. Without any loss of generality, subtracting the equation corresponding to the partial derivative with respect to $\beta_{L,1}$ from the rest of $N-1$ equations, it is easy to prove that $\beta_{L,1} = \beta_{L,2} = \cdots = \beta_{L,N} = \beta_{L}$(say) is one of the roots of this system of equations. Substituting this solution in one of the equations, we get
\begin{equation*}
\beta_L(N^2 h_t^2 h_e^2 \beta_L^4 (1+N B)\left(1+N B + A N P_s/\sigma^2\right)-1) = 0
\end{equation*}
This equation has the following two distinct and real solutions for the stationary points of the secrecy rate with respect to $\beta_L$: 
\begin{equation*}
\beta_{L,z} = 0, \mbox{ and } \beta_{L,glb}^2 = \frac{1}{N h_t h_e (1+N B) \sqrt{1+\frac{P_s}{\sigma^2}\frac{N A}{1+N B}}}
\end{equation*}

Now using the second derivative test, we can prove that when $h_t>h_e$, then $\beta_{L,z}$ and $\beta_{L,glb}$ are the points of global minimum and maximum, respectively. Similarly, for $h_e > h_t$, $\beta_{L,z}$ and $\beta_{L,glb}$ can be proved to be the points of global maximum and minimum, respectively. Given the convex nature of the secrecy rate function with respect to $\beta_L$ for $h_t>h_e$ and that $\beta_{L,max}$ is the largest value of the scaling factor $\beta_L$, we have the result
\begin{equation*}
\beta_{L,opt}^2 = \min(\beta_{L,max}^2, \beta_{L,glb}^2), \quad \mbox{if } h_t>h_e
\end{equation*}
\end{IEEEproof}

\begin{pavikl}
\label{lemma:ECGALReducedBeta2}
For ECGAL layered networks,
\begin{equation*}
(\bm{\beta_{1,opt}}, \ldots, \bm{\beta_{L-1,opt}}) = (\bm{\beta_{1,max}}, \ldots, \bm{\beta_{L-1,max}}),
\end{equation*}
where
\begin{align}
\beta_{1,1,max}^2 & \!\!=\cdots =\! \beta_{1,N,max}^2 \!\!= \!\beta_{1,max}^2\!\!= \frac{P}{P_{s} h_s^2 + \sigma^2},&\nonumber\\
\beta_{l,1,max}^2 & \!\! =\cdots =\!\beta_{l,N,max}^2 \!\!=\! \beta_{l,max}^2 \!\!= \frac{P}{P_{R_x,l}}, & \hspace{-1.1cm} l \!\in \!\{2, \!\cdots\!,\! L\!-\!1\} \label{eqn:beta_l_maxN}
\end{align}
with $P_{Rx,l} = P_s h_s^2 H_{1,l-1}^2 +\left[\sum_{j=1}^{l-1} G_{j,l-1}^2+1\right] \sigma^2$
\end{pavikl}
\begin{IEEEproof}
With optimum value of the scaling factors for the nodes in the last layer from Lemmas \ref{lemma:ECGALReducedBeta1}, the problem \eqref{eqn:eqProb} of computing the optimal network-wide scaling vector reduces to
\begin{align*}
\bm{\beta}_{opt} &= \argmax_{(\bm{\beta_{1}}, \ldots, \bm{\beta_{L-1}}, \bm{\beta_{L,opt}})} \frac{1+SNR_t}{1+SNR_e}.
\end{align*}
Similar to \cite[lemma 3]{116agrawalAgnihotri} for linear chain networks, we can show that $\frac{1+SNR_t}{1+SNR_e}$ is a quasi-convex function of $\bm{\beta}_{L-1}$ in the interval $[-\bm{\beta}_{L-1,max}, \bm{\beta}_{L-1,max}]$ for a given sub-vector $(\bm{\beta}_1, \ldots, \bm{\beta}_{L-2})$ of scaling factors of first $L-2$ relay layers and optimum sub-vector $\bm{\beta}_{L,opt}$ of the optimum scaling factors of the last relay layer. Thus, $\bm{\beta}_{L-1,opt} = \bm{\beta}_{L-1,max}$. Carrying out this process successively for relays in layer $L-2, \ldots, 1$, proves the lemma.
\end{IEEEproof}

Therefore, $\bm{\beta}_{opt}^N$ can be written as $\bm{\beta}_{opt}^N = ( \bm{\beta}_{1,max}^N, \cdots,  \bm{\beta}_{L-1,max}^N,  \bm{\beta}_{L,opt}^N)$. We use superscript $N$ to emphasize that the optimal scaling factors are computed for $N$ nodes in each layer.

With these optimum scaling factors, the SNR at the destination with $N$ relay nodes in each layer of a layered network with $L$ relay layers is given as follows
\begin{equation}
\label{eqn:snr_t_N}
S N R_{t,opt}^N = \frac{P_s h_s^2}{\sigma^2} \frac{\left(\prod_{l=1}^L (N \beta_l^N h_l)\right)^2}{1+ N\sum_{l=1}^{L}\left(\beta_l^N h_l \prod_{j=l+1}^{L} (N \beta_j^N h_j)\right)^2 }
\end{equation}
where 
\begin{align*}
\beta_l^N = \beta_{l,max}^N,   l=\{1, \cdots, L-1\}, \mbox{ and } \quad \beta_L^N = \beta_{L,opt}^N  
\end{align*}

Similarly, the SNR at the eavesdropper with $N$ relay nodes in each layer  is given as follows
\begin{equation}
\label{eqn:snr_e_N}
S N R_{e,opt}^N   =   \frac{{P_s h_s^2}\left(\prod_{l=1}^{L-1} (N \beta_l^N h_l)N \beta_L^N h_e\right)^2/{\sigma^2}}{1  +  N  \left( \beta_L^N h_e \right)^{ 2}  +  N \sum_{l=1}^{L-1} \left[ \beta_l^N  h_l   \left(\prod_{j=l+1}^{L-1} (N \beta_j^N  h_j) \right)   N  \beta_L^N  h_e \right]^2 }
\end{equation}

Now consider the network simplification scenario where only $k$ out of $N$ available relays in each layer are used. Using Lemma~\ref{lemma:ECGALReducedBeta1} and \ref{lemma:ECGALReducedBeta2}, we can solve secrecy rate optimization problem of \eqref{eqn:eqProb} and obtain the optimal solution $ \bm{\beta}_{opt}^k = \left(\beta_1^k, \cdots, \beta_L^k\right)$, where $\beta_l^k$ is the optimum scaling factor for all the nodes in layer $l$ and is given as follows:
\begin{align*}
\left(\beta_{l}^k\right)^2 &=\left(\beta_{l,max}^k\right)^2, \quad l=\{1, \cdots, L-1\}\\
\left(\beta_{L}^k\right)^2 &= \min\left\{\left(\beta_{L,max}^k\right)^2,\left(\beta_{L,glb}^k\right)^2\right\}
\end{align*}
where
\begin{align}
\left( \beta_{l,max}^k \right)^{ 2} & = \frac{P}{P_s h_s^2  \prod_{i=1}^{l-1}(k \beta_i^k h_i)^2  +   \left[ k  \sum_{i=1}^{l-1}  \left( \beta_i^k h_i    \prod_{j=i+1}^{l-1}    (k \beta_j^k h_j) \right)^{ 2}    + 1 \right]  \sigma^2},\label{eqn:beta_l_maxk}\\
\left(\beta_{L,glb}^k\right)^{ 2} &= \frac{1}{k h_L h_e \left(1  +  k B\right) \sqrt{1 +  \frac{P_s}{\sigma^2}\frac{k A}{1+k B}}} \label{eqn:betaLglb_k}
\end{align}
with $ A =  h_s^2\left( \prod_{l=1}^{L-1} k (\beta_l^k) h_l\right)^2,  \mbox{ and } B =  \!\left(\sum_{l=1}^{L-1} k (\beta_l^k)^2 h_l^2 \! \prod_{i=l+1}^{L-1}\! k^2 (\beta_i^k)^2 h_i^2\!\right)$

The corresponding optimal SNR at the destination and the eavesdropper is given as:
\begin{align}
S N R_{t,opt}^k &= \frac{P_s h_s^2}{\sigma^2} \frac{\left(\prod_{l=1}^L (k \beta_l^k h_l)\right)^2}{1+ k\sum_{l=1}^{L}\left(\beta_l^k h_l \prod_{j=l+1}^{L} (k \beta_j^k h_j)\right)^2 }\label{eqn:snr_t_k}\\
S N R_{e,opt}^k &  =   \frac{{P_s h_s^2}\left(\prod_{l=1}^{L-1} (k \beta_l^k h_l)k \beta_L^k h_e\right)^2/{\sigma^2}}{1  +  k  \left( \beta_L^k h_e \right)^{ 2}  +  k \sum_{l=1}^{L-1} \left[ \beta_l^k  h_l   \left(\prod_{j=l+1}^{L-1}     ( k \beta_j^k h_j ) \right)   k \beta_L^k h_e \right]^2 }\label{eqn:snr_e_k}
\end{align}

Let 
\begin{equation}
\label{eqn:lyrRs_N}
R_s^N = \frac{1}{2}\log\left(\frac{1+SNR_{t,opt}^N}{1+SNR_{e,opt}^N}\right) 
\end{equation}
and
\begin{equation}
\label{eqn:lyrRs_k}
R_s^k = \frac{1}{2}\log\left(\frac{1+SNR_{t,opt}^k}{1+SNR_{e,opt}^k}\right)
\end{equation} 
denote the optimal secure AF rate achieved by using all $N$ relays and any $k$ relays out of available $N$ relays in each layer of the ECGAL network, respectively. In the following we compute the additive gap $R_s^N - R_s^k$ for large $P_s$ and the multiplicative gap $R_s^N/R_s^k$ for small $P_s$.

Before discussing the upper-bounds on the additive and multiplicative gaps for ECGAL networks with arbitrary number of relay layers, consider the following example where we compute such bounds for an ECGAL network with two layers of relay nodes $(L=2)$ for any $N$ and $k$.

\begin{pavike}
\label{Ex:2lyrNetSim}
Consider an ECGAL network with two layers of relay nodes between the source and the destination, $L=2$. \\
\textbf{Case I: }$\beta_{2,opt}^2 = \beta_{2,max}^2$.\\
Using \eqref{eqn:snr_t_N} and \eqref{eqn:snr_e_N}, we have for this network
\begin{align}
SNR_{t,opt}^N &= \frac{P_s h_s^2}{\sigma^2} \frac{N^2 \beta_1^2 h_1^2 N^2 \left(\beta_2^N\right)^2 h_2^2}{1\!+\! N \!\left(\!\beta_2^N\!\right)^{\!2} \!\!h_2^2\! + \!N \beta_1^2 h_1^2 N^2\!\! \left(\!\beta_2^N\!\right)^{\!2}\!\! h_2^2} \label{eqn:snr_t_2lyr}\\
SNR_{e,opt}^N &= \frac{P_s h_s^2}{\sigma^2} \frac{N^2 \beta_1^2 h_1^2 N^2 \left(\beta_2^N\right)^2 h_e^2}{1\!+\! N \!\left(\!\beta_2^N\!\right)^{\!2} \!\!h_e^2\! + \!N \beta_1^2 h_1^2 N^2\!\! \left(\!\beta_2^N\!\right)^{\!2}\!\! h_e^2} \label{eqn:snr_e_2lyr}
\end{align}
with
\begin{align}
\beta_1^2                &= \beta_{1,max}^2     = \frac{P}{P_sh_s^2+\sigma^2} \label{eqn:beta1max}\\
\left(\!\beta_2^N\!\right)^{\!2} &= \!\left(\beta_{2,max}^N\right)^{\!2}\! = \frac{P}{P_sh_s^2 N^2 \beta_1^2 h_1^2 \!+\!\sigma^2\!\left(1\!+\! N \beta_1^2 h_1^2\right)} \label{eqn:beta2max}
\end{align}

Substituting for $\beta_1$ and $\left(\beta_2^N\right)$ in \eqref{eqn:snr_t_2lyr} and \eqref{eqn:snr_e_2lyr} and subsequently substituting the results in \eqref{eqn:lyrRs_N}, we get
\begin{equation}
R_s^N\! =\! \frac{1}{2} \log\!\left[\!\frac{1\!+\!\frac{\frac{P_{\!s} \!h_{\!s}^{\!2}\!}{\sigma^2} \frac{N^{\!2}\! P h_1^2}{\sigma^2}  \frac{N^{\!2}\! P h_2^2}{\sigma^2}}{\frac{P_{\!s} \!h_{\!s}^{\!2}\!}{\sigma^2} \left[\! \frac{N^{\!2}\! P h_1^2}{\sigma^2} +  \frac{ N P h_2^2}{\sigma^2} + 1 \!\right] + 1+ \frac{N  P h_1^2}{\sigma^2} + \frac{ N P h_2^2}{\sigma^2} + \frac{N P h_1^2}{\sigma^2} \frac{ N^{\!2}\! P h_2^2}{\sigma^2}}}{1\!+\!\frac{\frac{P_{\!s} \!h_{\!s}^{\!2}\!}{\sigma^2}  \frac{N^{\!2}\! P h_1^2}{\sigma^2}  \frac{N^{\!2}\! P h_e^2}{\sigma^2}}{\frac{P_{\!s} \!h_{\!s}^{\!2}\!}{\sigma^2} \left[\! \frac{N^{\!2}\! P h_1^2}{\sigma^2} +  \frac{N P h_e^2}{\sigma^2} + 1 \!\right] + 1+  \frac{N P h_1^2}{\sigma^2} +  \frac{N P h_e^2}{\sigma^2} + \frac{N P h_1^2}{\sigma^2} \frac{N^{\!2}\!  P h_e^2}{\sigma^2}}}\right]
\end{equation}

Similarly,
\begin{equation}
R_s^k\! =\! \frac{1}{2} \log\!\left[\!\frac{1\!+\!\frac{\frac{P_{\!s} h_s^2}{\sigma^2} k^2 \frac{P h_1^2}{\sigma^2} k^2 \frac{P h_2^2}{\sigma^2}}{\frac{P_{\!s} h_s^2}{\sigma^2} \left[\!k^2 \frac{P h_1^2}{\sigma^2} + k \frac{P h_2^2}{\sigma^2} + 1 \!\right] + 1+ k \frac{P h_1^2}{\sigma^2} + k \frac{P h_2^2}{\sigma^2} +k \frac{P h_1^2}{\sigma^2} k^2 \frac{P h_2^2}{\sigma^2}}}{1\!+\!\frac{\frac{P_{\!s} h_s^2}{\sigma^2} k^2 \frac{P h_1^2}{\sigma^2} k^2 \frac{P h_e^2}{\sigma^2}}{\frac{P_{\!s} h_s^2}{\sigma^2} \left[\!k^2 \frac{P h_1^2}{\sigma^2} + k \frac{P h_e^2}{\sigma^2} + 1 \!\right] + 1+ k \frac{P h_1^2}{\sigma^2} + k \frac{P h_e^2}{\sigma^2} +k \frac{P h_1^2}{\sigma^2} k^2 \frac{P h_e^2}{\sigma^2}}}\right]
\end{equation}

Thus, we have
\begin{align}
\lim_{P_s \rightarrow \infty} R_s^N  - R_s ^k & =  \frac{1}{2}\log \!\left[\!\frac{\!1+ \frac{N^2 P h_2^2/\sigma^2}{1\!+ \frac{h_2^2}{N h_1^2}+ \frac{\sigma^2}{N^2 P h_1^2 }}}{1\!+ \frac{k^2 P h_2^2/\sigma^2}{1\!+ \frac{h_2^2}{k h_1^2}+ \frac{\sigma^2}{k^2 P h_1^2 } }} \ \frac{1\!+ \frac{k^2 P h_e^2/\sigma^2}{1+ \frac{h_e^2}{k h_1^2}+ \frac{\sigma^2}{k^2 P h_1^2 } }}{1\!+ \frac{N^2 P h_e^2/\sigma^2}{1\!+ \frac{h_e^2}{N h_1^2}+ \frac{\sigma^2}{N^2 P h_1^2 } }}\!\right]\nonumber\\
			&\leq \frac{1}{2}\log\! \left[1\!+\!\frac{h_2^2}{h_1^2}\!\left(\!\frac{1}{k}\!-\!\frac{1}{N}\!\right)\!\!+\! \frac{\sigma^2}{P h_1^2}\! \left(\!\frac{1}{k^2}\!-\!\frac{1}{N^2}\!\right)\!\right] \nonumber\\
	&\quad + \frac{1}{2}\log\! \left[1\!+\!\frac{\sigma^2}{P h_e^2}\!\left(\!\frac{1}{k^2}\!-\!\frac{1}{N^2}\!\right)\!\!+\! \frac{\sigma^2}{P h_1^2} \!\left(\!\frac{1}{k^3}\!-\!\frac{1}{N^3}\!\right)\!\! +\! \frac{\sigma^4}{P h_1^2 P h_e^2}\! \left(\!\frac{1}{k^4}\!-\!\frac{1}{N^4}\!\right)\right]
\end{align}
and
\begin{align}
\lim_{P_s \rightarrow 0} \frac{R_s^N}{R_s^k}  &= \lim_{P_s \rightarrow 0}\log\left. \left[  \frac{1 + \frac{N P_s h_s^2/\sigma^2}{1+\frac{\sigma^2}{N^{ 2}  P} \left[ \frac{1}{h_{ 1}^{ 2}}  +  \frac{1}{h_{ 2}^{ 2}} +  \frac{ \sigma^2}{ N  h_{ 1}^{ 2}  h_{ 2}^{ 2}  P} \right]}}{1 + \frac{N P_s h_s^2/\sigma^2}{1+\frac{\sigma^2}{N^{ 2}  P} \left[ \frac{1}{h_{ 1}^{ 2}}  +  \frac{1}{h_{ e}^{ 2}} +  \frac{ \sigma^2}{ N  h_{ 1}^{ 2}  h_{ e}^{ 2}  P} \right]}} \right]  \middle/   \log \left[  \frac{1 + \frac{k P_s h_s^2/\sigma^2}{1+\frac{\sigma^2}{k^{ 2}  P} \left[ \frac{1}{h_{ 1}^{ 2}}  +  \frac{1}{h_{ 2}^{ 2}} +  \frac{ \sigma^2}{ k  h_{ 1}^{ 2}  h_{ 2}^{ 2}  P} \right]}}{1 + \frac{k P_s h_s^2/\sigma^2}{1+\frac{\sigma^2}{k^{ 2}  P} \left[ \frac{1}{h_{ 1}^{ 2}}  +  \frac{1}{h_{ e}^{ 2}} +  \frac{ \sigma^2}{ k  h_{ 1}^{ 2}  h_{ e}^{ 2}  P} \right]}} \right]\right.\nonumber\\
	&\leq \left(\!\!\frac{N}{k}\!\!\right)^{\!4}\! \left(\!\frac{1\!+\!N\frac{P h_1^2}{\sigma^2}}{1\!+\!k\frac{P h_1^2}{\sigma^2}}\!\right) \left(\!\frac{ 1\!+\! k\frac{P h_1^2}{\sigma^2} \!+\! k\frac{P h_e^2}{\sigma^2} \!+\! k^3 \frac{P h_1^2}{\sigma^2} \frac{P h_e^2}{\sigma^2}}{1\!+\! N\frac{P h_1^2}{\sigma^2} \!+\! N\frac{P h_2^2}{\sigma^2} \!+\! N^3 \frac{P h_1^2}{\sigma^2} \frac{P h_2^2}{\sigma^2}}\!\right)\left(\!\frac{1\!+\! k\frac{P h_1^2}{\sigma^2} \!+\! k\frac{P h_2^2}{\sigma^2} \!+\! k^3 \frac{P h_1^2}{\sigma^2} \frac{P h_2^2}{\sigma^2}}{1\!+\! N\frac{P h_1^2}{\sigma^2} \!+\! N\frac{P h_e^2}{\sigma^2} \!+\! N^3 \frac{P h_1^2}{\sigma^2} \frac{P h_e^2}{\sigma^2}}\!\right)\nonumber\\
	&\leq \left(\!\frac{N}{k}\!\right)^{\!2} \left[1\!+\!\frac{\sigma^2}{ P h_1^2}\!\left(\frac{1}{k^2}\!-\!\frac{1}{N^2}\right)\! +\! \frac{\sigma^2}{ P}\!\left(\!\frac{1}{k^2 h_e^2}\!-\!\frac{1}{N^2 h_2^2}\!\right)\! +\! \frac{\sigma^2}{ P h_1^2}\frac{\sigma^2}{ P}\!\left(\!\frac{1}{k^3 h_e^2}\!-\!\frac{1}{N^3 h_2^2}\!\right)\!\right]
\end{align}
\textbf{Case II: }$\beta_{2,opt}^2 = \beta_{2,glb}^2$.\\
From Lemma~\ref{lemma:ECGALReducedBeta1}, we have
\begin{align}
\left(\beta_{2,glb}^N\right)^2 &= \frac{1}{N h_2 h_e \!\left(1 \!+\! N^2 \beta_1^2 h_1^2\right)\sqrt{1\!+\!\frac{P_s h_s^2}{\sigma^2} \frac{N^3 \beta_1^2 h_1^2}{1+ N^2 \beta_1^2 h_1^2}}} \label{eqn:beta2glbN}
\end{align}

Substituting for $\beta_1$ and $\left(\beta_2^N\right)$ in \eqref{eqn:snr_t_2lyr} and \eqref{eqn:snr_e_2lyr} and subsequently substituting the results in \eqref{eqn:lyrRs_N}, we get
\begin{equation}
R_s^N = \frac{1}{2} \log\!\left[\!\frac{1+\frac{N P_s h_s^2/\sigma^2}{1+\frac{\sigma^2}{N^2 P h_1^2}\left[1+\frac{P_s h_s^2 }{\sigma^2} + \frac{ h_e}{ h_2} D\sqrt{1+\frac{P_s h_s^2}{\sigma^2}\frac{N^3 P h_1^2}{\sigma^2}D}\right]}}{1+\frac{N P_s h_s^2/\sigma^2}{1+\frac{\sigma^2}{N^2 P h_1^2}\left[1+\frac{P_s h_s^2 }{\sigma^2} + \frac{ h_2}{ h_e} D\sqrt{1+\frac{P_s h_s^2}{\sigma^2}\frac{N^3 P h_1^2}{\sigma^2}D}\right]}}\!\right]
\end{equation}
with $D={1+\frac{P_s h_s^2}{\sigma^2} + \frac{N^2 P h_1^2}{\sigma^2}}$.

Similarly,
\begin{equation}
R_s^k = \frac{1}{2} \log\!\left[\!\frac{1+\frac{k P_s h_s^2/\sigma^2}{1+\frac{\sigma^2}{k^2 P h_1^2}\left[1+\frac{P_s h_s^2 }{\sigma^2} + \frac{ h_e}{ h_2} D'\sqrt{1+\frac{P_s h_s^2}{\sigma^2}\frac{k^3 P h_1^2}{\sigma^2}D'}\right]}}{1+\frac{k P_s h_s^2/\sigma^2}{1+\frac{\sigma^2}{k^2 P h_1^2}\left[1+\frac{P_s h_s^2 }{\sigma^2} + \frac{ h_2}{ h_e} D'\sqrt{1+\frac{P_s h_s^2}{\sigma^2}\frac{k^3 P h_1^2}{\sigma^2}D'}\right]}}\!\right]
\end{equation}
with $D'={1+\frac{P_s h_s^2}{\sigma^2} + \frac{k^2 P h_1^2}{\sigma^2}}$.

Thus, we have
\begin{align}
\lim_{P_s \rightarrow \infty} R_s^N - R_s ^k &= \frac{1}{2}\log \left[\frac{1+\frac{N^3 Ph_1^2 /\sigma^2}{1 + \frac{h_e }{h_2}\sqrt{\!1\!+\! \frac{N^3 P h_1^2}{\sigma^2 }}}}{1+\frac{k^3 P h_1^2/\sigma^2}{1 + \frac{h_e }{ h_2}\sqrt{\!1\!+\! \frac{k^3 P h_1^2}{\sigma^2 }}}} \frac{1+\frac{k^3 P h_1^2/\sigma^2}{1+ \frac{h_2 }{h_e}\sqrt{\!1\!+\! \frac{k^3 P h_1^2}{\sigma^2 }}}}{1+\frac{N^3 Ph_1^2 /\sigma^2}{1 + \frac{h_2 }{ h_e}\sqrt{\!1\!+\! \frac{N^3 P h_1^2}{\sigma^2 }}}}\right] \nonumber\\	
											   &\leq \frac{1}{2}\log \left[\left(\frac{N}{k}\right)^3 \frac{1 + \frac{h_e}{  h_2} \sqrt{\!1\!+\! \frac{k^3 P h_1^2}{\sigma^2 }}}{1+ \frac{h_e }{  h_2}\sqrt{\!1\!+\! \frac{N^3 P h_1^2}{\sigma^2 }}}\ \frac{1 \!+\! \frac{h_2 }{    h_e}\sqrt{\!1\!+\!\frac {k^3 P h_1^2}{\sigma^2 }}\!+\!\frac{k^3 Ph_1^2 }{\sigma^2}}{1\! +\! \frac{h_2 }{  h_e}\sqrt{\!1\!+\!\frac {N^3 P h_1^2}{\sigma^2 }}\!+\!\frac{N^3 Ph_1^2}{\sigma^2}}\ \frac{1\! +\! \frac{h_2 }{  h_e}\sqrt{\!1\!+\! \frac{N^3 P h_1^2}{\sigma^2 }}}{1\!+\! \frac{h_2 }{ h_e}\sqrt{\!1\!+\! \frac{k^3 P h_1^2}{\sigma^2 }}}\right] \nonumber\\
											   & \leq \frac{3}{4} \log\!\left(\frac{N}{k}\right) \!+\! \frac{1}{2}\!\log\! \left[1\!+\! \frac{\sigma^2}{P h_1^2}\! \left[\left\{\frac{1}{k^3}\!-\!\frac{1}{N^3}\right\}+\!\frac{h_2}{h_e}\!\left\{\!\sqrt{\frac{1}{k^6}\!+\!\frac{P h_1^2}{k^3 \sigma^2}}-\!\sqrt{\frac{1}{N^6}\!+\!\frac{P h_1^2}{N^3 \sigma^2}}\right\}\right]\right]
\end{align}
and
\begin{align}
\lim_{P_s \rightarrow 0} \frac{R_s^N}{R_s^K} &=\lim_{P_s \rightarrow 0} \log \left( \left.\frac{1+\frac{N P_s h_s^2/\sigma^2}{1+\frac{\sigma^2}{N^2 P}\left(\frac{ 1}{ h_1^2} +\frac{h_e}{h_1^2 h_2} + \frac{N^2 P h_e}{\sigma^2 h_2}\right)}}{1+\frac{N P_s h_s^2/\sigma^2}{1+\frac{\sigma^2}{N^2 P}\left(\frac{ 1}{ h_1^2} +\frac{h_2}{h_1^2 h_e} + \frac{N^2 P h_2}{\sigma^2 h_e}\right)}}\right)\middle/\right.\log \left( \frac{1+\frac{K P_s h_s^2/\sigma^2}{1+\frac{\sigma^2}{K^2 P}\left(\frac{ 1}{ h_1^2} +\frac{h_e}{h_1^2 h_2} + \frac{K^2 P h_e}{\sigma^2 h_2}\right)}}{1+\frac{K P_s h_s^2/\sigma^2}{1+\frac{\sigma^2}{K^2 P}\left(\frac{ 1}{ h_1^2} +\frac{h_2}{h_1^2 h_e} + \frac{K^2 P h_2}{\sigma^2 h_e}\right)}} \right)\nonumber\\
		&\leq \lim_{P_s \rightarrow 0} \!\left(\!\frac{N}{k}\!\right)^{\!3}\! \frac{\left[\!1\!+\!k^2\frac{P h_1^2}{\sigma^2}\right](h_2\!+\!h_e) \!+\! k^3 \frac{P_s h_s^2}{\sigma^2}\frac{P h_1^2}{\sigma^2} h_2}{\left[\!1\!+\!N^2\frac{P h_1^2}{\sigma^2}\right](h_2\!+\!h_e) \!+\! N^3 \frac{P_s h_s^2}{\sigma^2}\frac{P h_1^2}{\sigma^2} h_e} \nonumber\\
		& \leq\! \max\left\{\left(\frac{N}{k}\right)^3 \frac{\left(1+k^2\frac{P h_1^2}{\sigma^2}\right) }{\left(1+N^2\frac{P h_1^2}{\sigma^2}\right) }, \frac{h_2}{h_e} \right\}\nonumber\\
		&\leq \max\left\{\!\left(\frac{N}{k}\right) \left[1+\frac{\sigma^2}{P h_1^2}\left(\frac{1}{k^2}-\frac{1}{N^2}\right)\right], \frac{h_2}{h_e} \right\}
\end{align}
for arbitrary $N$ and $k$ relays in each layer.\hfill $\blacksquare$
\end{pavike}

However, for ECGAL networks with arbitrary number of relay layers, it is analytically hard to compute such upper-bounds on the additive and multiplicative gaps between the optimal end-to-end performances with and without network simplification for any $N$ and $k$. Therefore, in the following we attempt to analyze the scaling behavior of such upper bounds with large $N$ and $k$.\\
\textbf{Case I:} $\beta_{L,opt}^2 = \beta_{L,max}^2$.

Using \eqref{eqn:beta_l_maxN} and \eqref{eqn:beta_l_maxk}, we have for large $N$ and $k$:
\begin{align}
(\text{for } P_s \!\rightarrow \!\infty)  & \prod\limits_{i=1}^{l} \left(\beta_i^N\right)^2\! = \! \frac{P/(P_s h_s^2)}{N^{2(l-1)} \prod\limits_{i=1}^{l-1} h_i^2},  1\leq l \leq L \label{eqn:betasProdPsInftyN}\\
                                      & \prod\limits_{i=1}^{l} \left(\beta_i^k\right)^2 \! = \! \frac{P/(P_s h_s^2)}{k^{2(l-1)} \prod\limits_{i=1}^{l-1} h_i^2},  1\leq l \leq L  \label{eqn:betasProdPsInftyk}\\
(\text{for } P_s \!\rightarrow\! 0)       & \prod\limits_{i=1}^{l} \left(\beta_i^N\right)^{\!2} \!=\! \frac{P/\sigma^2}{N^{2(l-1\!)\!-1} \!\prod\limits_{i=1}^{l-1}\! h_i^2}, 1\leq l \leq L \label{eqn:betasProdPsZeroN}\\
                                      & \prod\limits_{i=1}^{l} \left(\beta_i^k\right)^2 \!=\! \frac{P/\sigma^2}{k^{2(l-1)-1} \!\prod\limits_{i=1}^{l-1}\! h_i^2},  1\leq l \leq L  \label{eqn:betasProdPsZerok}
\end{align}
Then from \eqref{eqn:snr_t_N}, \eqref{eqn:snr_e_N} and \eqref{eqn:betasProdPsInftyN} we obtain for large $N$ and $P_s \rightarrow \infty$:
\begin{align}
S\!N\!R_t^N \sim  \frac{N^2 P h_L^2}{\sigma^2 \left(1 \!+\!\frac{a}{N} \right)}, \text{ and } S\!N\!R_e^N \sim \frac{N^2 P h_e^2}{\sigma^2 \left(1\! +\!\frac{h_e^2}{h_L^2} \frac{a}{N}\right)} \label{eqn:snr_N_Ps_infty}
\end{align}
where, $ a= h_L^2 \sum_{l=1}^{L-1}\frac{1}{h_{l}^2}$

Similarly, from \eqref{eqn:snr_t_k}, \eqref{eqn:snr_e_k} and \eqref{eqn:betasProdPsInftyk} we obtain for large $k$ and $P_s \rightarrow \infty$:
\begin{align}
S\!N\!R_t^k \sim  \frac{k^2 P h_L^2}{\sigma^2 \left(1 +\frac{a}{k} \right)},\text{ and }S\!N\!R_e^k \sim \frac{k^2 P h_e^2}{\sigma^2 \left(1 +\frac{h_e^2}{h_L^2} \frac{a}{k}\right)} \label{eqn:snr_k_Ps_infty}
\end{align}
And for large $N$ and $k$, and $P_s \rightarrow 0$, from \eqref{eqn:snr_t_N}, \eqref{eqn:snr_e_N} and \eqref{eqn:betasProdPsZeroN} we obtain
\begin{align}
\label{eqn:snr_N_Ps_0}
S\!N\!R_t^N \sim \frac{N P_s h_s^2/\sigma^2}{1\!+\!\frac{\sigma^2}{N^2 P h_L^2}\!\left[\!1\!+\!\frac{h_L^2}{h_1^2}\!+\!\frac{b}{N}\right]} \mbox{, and } S\!N\!R_e^N \sim \frac{N P_s h_s^2/\sigma^2}{1\!+\!\frac{\sigma^2}{N^2 P}\!\left[\frac{h_L^2}{h_e^2}\!+\!\frac{h_L^2}{h_1^2}\!+\!\frac{b}{N}\right]} 
\end{align}
and from \eqref{eqn:snr_t_k}, \eqref{eqn:snr_e_k} and \eqref{eqn:betasProdPsZerok}
\begin{align}
\label{eqn:snr_k_Ps_0}
S\!N\!R_t^k \sim \frac{k P_s h_s^2/\sigma^2}{1\!+\!\frac{\sigma^2}{k^2 P h_L^2}\!\left[\!1\!+\!\frac{h_L^2}{h_1^2}\!+\!\frac{b}{k}\right]}  \mbox{, and } S\!N\!R_e^k \sim \frac{k P_s h_s^2/\sigma^2}{1\!+\!\frac{\sigma^2}{k^2 P}\!\left[\frac{h_L^2}{h_e^2}\!+\!\frac{h_L^2}{h_1^2}\!+\!\frac{b}{k}\right]} 
\end{align}
where, $b =h_L^2\sum_{i=2}^{L-1}\frac{1}{h_i^2}$.

With these results we are now ready to compute the upper bound on additive and multiplicative gap between the optimal performance of AF relaying with and without network simplification.

First, we consider the upper bound on the additive gap. Substituting the SNR values from \eqref{eqn:snr_N_Ps_infty} and \eqref{eqn:snr_k_Ps_infty} in \eqref{eqn:lyrRs_N} and \eqref{eqn:lyrRs_k} respectively, we obtain the following upper bound on the additive gap $\bar{R}_s^N - \bar{R}_s^k$ for $P_s \rightarrow \infty$ and asymptotically large $N$ and $k$ satisfying $N,k = o(P_s)$:
\begin{equation}
\bar{R}_s^N \!\!-\! \bar{R}_s^k \leq  \frac{1}{2}\!\log \!\left[1\!+\! {a \left(\frac{1}{k}\!-\!\frac{1}{N}\right)} \right] +\!\frac{1}{2}\!\log\!\left[1\!+\! a {\frac{\sigma^2}{P h_L^2}\!\left(\frac{1}{k^3}\!-\!\frac{1}{N^3}\right)\!\! +\! \frac{\sigma^2}{P h_e^2}\! \left(\frac{1}{k^2}\!-\!\frac{1}{N^2}\right)}\right]\label{eqn:addGapLlyrBetamax}
\end{equation}

Next, we consider the upper bound on the multiplicative gap. Substituting the SNR values from \eqref{eqn:snr_N_Ps_0} and \eqref{eqn:snr_k_Ps_0} in \eqref{eqn:Rs_N} and \eqref{eqn:Rs_k} resp., we obtain the following upper bound on the multiplicative gap $\bar{R}_s^N / \bar{R}_s^k$ for $P_s \rightarrow 0$ and asymptotically large $N$ and $k$ satisfying $N,k = o(1/P_s)$:
\begin{align}
\label{eqn:mulGapLlyrBetamax}
\frac{\bar{R}_s^N}{\bar{R}_s^k}   \leq & \left(\!\frac{N}{k}\!\right)^{\!2}\left[1\!+\!\frac{\sigma^2}{ P h_1^2}\left(\frac{1}{k^2}-\frac{1}{N^2}\right)+\frac{\sigma^2}{P}\!\left(\frac{1}{k^2 h_e^2}-\frac{1}{N^2 h_L^2}\right)\! +\! \frac{ \sigma^2}{P h_L^2}\left(\frac{1}{k^3}\!-\!\frac{1}{N^3}\right)b\right]
\end{align}
\textbf{Case II: }$\beta_{L,opt}^2 = \beta_{L,glb}^2$.

From \eqref{eqn:betaLglb} and \eqref{eqn:betasProdPsInftyN} we have, for $P_s \rightarrow \infty$ and asymptotically large $N$ : 
\begin{equation}
\left(\beta_{L,glb}^N\right)^2 \!= \frac{1}{N h_L h_e \left[1\! +\!{h_{L-1}^2} \! \sum_{l=1}^{L-2}{1}/{h_{l}^2}\right]\! \sqrt{1\!+\! \frac{ N^3 P h_{L\!-\!1}^2/\sigma^2 }{ 1+{h_{L\!-\!1}^2}\! \sum_{l=1}^{L-2}{1}/{h_{l}^2}}}} \label{eqn:betaLglbNInfty}
\end{equation}

Thus, from \eqref{eqn:snr_t_N} and \eqref{eqn:snr_e_N} we have, for $P_s \rightarrow \infty$ and asymptotically large $N$ : 
\begin{align}
S\!N\!R_{t,opt}^N \! \sim \frac{ N^3 P  h_L^2/ \sigma^2}{N^{\frac{3}{2}} h_e \sqrt{a\!+\!{aP}/{\sigma^2}}},&&  S\!N\!R_{e,opt}^N \! \sim \frac{ k^3 P  h_L^2/ \sigma^2}{k^{\frac{3}{2}} h_e \sqrt{a\!+\!{aP}/{\sigma^2}}} \label{eqn:snr_N_Ps_infty_glb}
\end{align}

Similarly, for $P_s \rightarrow \infty$ and asymptotically large $k$ we have: 
\begin{align}
S\!N\!R_{t,opt}^k  \sim \frac{ k^3 P  h_L^2/ \sigma^2}{k^{\frac{3}{2}} h_e \sqrt{a\!+\!{aP}/{\sigma^2}}}, && S\!N\!R_{e,opt}^k  \sim \frac{ k^3 P  h_L^2/ \sigma^2}{k^{\frac{3}{2}} h_e \sqrt{a\!+\!{aP}/{\sigma^2}}} \label{eqn:snr_k_Ps_infty_glb}
\end{align}

Substituting these SNR values from \eqref{eqn:snr_N_Ps_infty_glb} and \eqref{eqn:snr_k_Ps_infty_glb} in \eqref{eqn:Rs_N} and \eqref{eqn:Rs_k} respectively, we obtain the following upper bound on additive gap $\bar{R}_s^N - \bar{R}_s^k$ for $P_s \rightarrow \infty$:
\begin{align}
\bar{R}_s^N - \bar{R}_s^k \leq  &\frac{3}{4}\log \!\left(\frac{N}{K}\right) \!+\! \frac{1}{2}\log\!\left[\!1\!+\! \sqrt{\frac{\sigma^2 a}{P h_e^2}}\! \left(\frac{1}{K^{\frac{3}{2}}}\!-\!\frac{1}{N^{\frac{3}{2}}}\right) \!+\! \frac{\sigma^2 a}{ P h_L^2}\! \left(\frac{1}{K^3}\!-\!\frac{1}{N^3}\right)\right]\label{eqn:addGapLlyrBetaglb}
\end{align}

Now, we consider the upper bound on the multiplicative gap for $P_s \rightarrow 0$. From \eqref{eqn:betaLglb} and \eqref{eqn:betasProdPsZeroN} we have, for $P_s \rightarrow 0$ and asymptotically large $N$ : 
\begin{align}
\left(\beta_{L,glb}^N\right)^2 &= \frac{1}{N h_L h_e h_{L-1}^2 \left( \frac{N^3 P}{\sigma^2}+\frac{N}{h_1^2} +\sum_{i=2}^{L-1}\frac{1}{h_{i}^2} \right)} \label{eqn:betaLglbNPs0}
\end{align}
Thus, from \eqref{eqn:snr_t_N} and \eqref{eqn:snr_e_N} we have, for $ P_s \rightarrow 0 $ and asymptotically large $ N $ : 
\begin{align}
\label{eqn:snr_N_Ps_0_glb}
SNR_{t,opt}^N  \sim  \frac{P_s h_s^2}{\sigma^2} \frac{ \frac{N^4 {P h_L^2}/{\sigma^2}}{ h_L h_e \left( \frac{N^3 P}{\sigma^2}+\frac{N}{h_1^2} +\sum_{i=2}^{L-1}\frac{1}{h_{i}^2} \right)}}{1+\frac{h_L}{h_e}},\mbox{ and  } SNR_{e,opt}^N  \sim & \frac{P_s h_s^2}{\sigma^2} \frac{ \frac{N^4 {P h_e^2}/{\sigma^2}}{ h_L h_e \left( \frac{N^3 P}{\sigma^2}+\frac{N}{h_1^2} +\sum_{i=2}^{L-1}\frac{1}{h_{i}^2} \right)}}{1+\frac{h_e}{h_L}} 
\end{align}

Similarly, for $P_s \rightarrow 0$ and asymptotically large $k$ we have: 
\begin{align}
\label{eqn:snr_k_Ps_0_glb}
SNR_{t,opt}^k  \sim   \frac{P_s h_s^2}{\sigma^2} \frac{ \frac{k^4 {P h_L^2}/{\sigma^2}}{ h_L h_e \left( \frac{k^3 P}{\sigma^2}+\frac{k}{h_1^2} +\sum_{i=2}^{L-1}\frac{1}{h_{i}^2} \right)}}{1+\frac{h_L}{h_e}},\mbox{ and } SNR_{e,opt}^k  \sim  & \frac{P_s h_s^2}{\sigma^2} \frac{ \frac{k^4 {P h_e^2}/{\sigma^2}}{ h_L h_e \left( \frac{k^3 P}{\sigma^2}+\frac{k}{h_1^2} +\sum_{i=2}^{L-1}\frac{1}{h_{i}^2} \right)}}{1+\frac{h_e}{h_L}} 
\end{align}

Substituting these SNR values from \eqref{eqn:snr_N_Ps_0_glb} and \eqref{eqn:snr_k_Ps_0_glb} in \eqref{eqn:lyrRs_N} and \eqref{eqn:lyrRs_k} respectively, we obtain the following upper bound on multiplicative gap $\bar{R}_s^N / \bar{R}_s^k$ for $P_s \rightarrow 0$:
\begin{equation}
\label{eqn:mulGapLlyrBetaglb}
\frac{\bar{R}_s^N}{\bar{R}_s^k} \leq \max \left\{ \left(\frac{N}{k}\right) \left[1 \!+\! \frac{\sigma^2}{P h_1^2}\left(\frac{1}{k^2}\!-\!\frac{1}{N^2}\right) \!+\! \frac{\sigma^2 b}{P}\left(\frac{1}{k^3}\!-\!\frac{1}{N^3}\right)\right], \frac{h_L}{h_e} \right\}
\end{equation}
The results on the asymptotic behaviour of additive and multiplicative gaps are summarized in the following lemma:
\begin{pavikl}
For ECGAL network, the asymptotic additive and multiplicative gaps between the optimal performance of Amplify-and-Forward relaying obtained in terms of maximum achievable secrecy rate with and without network simplification are bounded from above as:

For $\beta_{L,opt}=\beta_{L,max}$,
\begin{align*}
\bar{R}_s^N \!\!-\! \bar{R}_s^k \leq & \frac{1}{2}\!\log \!\left[1\!+\! {a \left(\frac{1}{k}\!-\!\frac{1}{N}\right)} \right]\! +\!\frac{1}{2}\!\log\!\left(1\!+\! a {\frac{\sigma^2}{P h_L^2}\!\left(\frac{1}{k^3}\!-\!\frac{1}{N^3}\right)\!\! +\! \frac{\sigma^2}{P h_e^2}\! \left(\frac{1}{k^2}\!-\!\frac{1}{N^2}\right)}\right]\\
						& \leq \frac{1}{2}\!\log \!\left[1\!+\! a \right]+\!\frac{1}{2}\!\log\!\left[1\!+\! a {\frac{\sigma^2}{P h_L^2}\!\! +\! \frac{\sigma^2}{P h_e^2}}\right]\\
\frac{\bar{R}_s^N}{\bar{R}_s^k}   \leq & \left(\!\frac{N}{k}\!\right)^{\!2}\left[1\!+\!\frac{\sigma^2}{ P h_1^2}\left(\frac{1}{k^2}-\frac{1}{N^2}\right)+\frac{\sigma^2}{P}\!\left(\frac{1}{k^2 h_e^2}-\frac{1}{N^2 h_L^2}\right)\! +\! \frac{ \sigma^2}{P h_L^2}\left(\frac{1}{k^3}\!-\!\frac{1}{N^3}\right)b\right]
\end{align*}

and, for $\beta_{L,opt}=\beta_{L,glb}$,
\begin{align*}
\bar{R}_s^N - \bar{R}_s^k \leq  &\frac{3}{4}\log \!\left(\frac{N}{K}\right) \!+\! \frac{1}{2}\log\!\left[\!1\!+\! \sqrt{\frac{\sigma^2 a}{P h_e^2}}\! \left(\frac{1}{K^{\frac{3}{2}}}\!-\!\frac{1}{N^{\frac{3}{2}}}\right) \!+\! \frac{\sigma^2 a}{ P h_L^2}\! \left(\frac{1}{K^3}\!-\!\frac{1}{N^3}\right)\right]\\
\frac{\bar{R}_s^N}{\bar{R}_s^k} \leq &\max \left\{ \left(\frac{N}{k}\right) \left[1 \!+\! \frac{\sigma^2}{P h_1^2}\left(\frac{1}{k^2}\!-\!\frac{1}{N^2}\right) \!+\! \frac{\sigma^2 b}{P}\left(\frac{1}{k^3}\!-\!\frac{1}{N^3}\right)\right], \frac{h_L}{h_e} \right\}
\end{align*}

with $a=h_L^2\sum_{i=1}^{L-1}\frac{1}{h_i^2}$ and $b=h_L^2\sum_{i=2}^{L-1}\frac{1}{h_i^2}.$
\end{pavikl}

\noindent \textit{Discussion:} The results in this lemma show that asymptotically (in source power), for the case where the constraint on scaling factors of the nodes is satisfied with strict equality $\beta_{L,opt}=\beta_{L,max}$, the additive gap is independent of the ratio $N/k$ and increases at most logarithmically with $L$ and the corresponding multiplicative gap increases at most quadratically with ratio $N/k$ and $L$. Similarly, when the constraint on scaling factors of the nodes which eavesdropper snoops on is satisfied with strict inequality, $\beta_{L,opt}=\beta_{L,glb}$, the additive gap increases at most logarithmically with ratio $N/k$ and $L$, and the corresponding multiplicative gap increases at most linearly with ratio $N/k$ and $L$. 

\section{Conclusion and Future Work}
\label{sec:cnclsn}
Exact characterization of the optimum secure AF rate in general layered relay networks is an important but computationally intractable problem. We take an approach based on the notion of network simplification to approximate the optimal secure AF rate within small additive and multiplicative gaps in the symmetric Gaussian N-relay diamond network and a class of symmetric layered networks while simultaneously reducing the computational effort of solving this problem. To the best of our knowledge, this work provides the first characterization of
the performance of network simplification in AF relay networks in the presence of an eavesdropper. In future, we plan to extend this work to general layered networks.

\ifCLASSOPTIONcaptionsoff
  \newpage
\fi

\bibliographystyle{IEEEtran}
{\normalsize
\bibliography{sec_net_simp_bib}
}

\end{document}